\documentstyle[11pt,epsf]{article}

\begin{document}

\baselineskip 0.8cm

\title{ \Large\bf
Chiral symmetry breaking, the superspace gap equation and\\
 the no-renormalization theorem}
\author{
Andras Kaiser\thanks{E-mail: andras.kaiser@yale.edu}\quad and
 Stephen B. Selipsky\thanks{E-mail: stephen@genesis1.physics.yale.edu ;
 address after September 1, 1997: Dept.\ of Physics, Washington University,
 Campus Box 1105, Saint Louis, MO 63110-4899}
\\ \\ {\it
 Department of Physics, Yale University, New Haven, CT 06520-8120}}

\date{August 15, 1997}

\maketitle

\begin{picture}(0,0)(0,0)
\put(365,245){hep-th/9708087}
\put(365,230){YCTP-P14-97}
\end{picture}
\vspace{-24pt}

\begin{abstract}
   Solutions of the superspace Schwinger-Dyson equations, describing mass
generation and chiral symmetry breaking in supersymmetric gauge theory,
need not be constrained to vanish by no-renormalization theorems,
nor by special choices of gauge parameter.  Thus symmetry breaking
vacuum structures remain possible (as in non-supersymmetric gauge theory),
inviting comparison with predictions of an alternative approach
based on holomorphy and the Wilsonian effective action.
\end{abstract}

   The gap equation formalism has frequently been employed to describe
the vacuum structure of strongly interacting gauge theories.
Chiral symmetry breaking provides an especially interesting application.
The resulting conclusions can now be compared with those of independent
approaches based on holomorphy and the Wilsonian effective action
\cite{holomorf}, in the special case of supersymmetric gauge theories.
Some previous supersymmetric gap equation analyses \cite{Shamir,ANS}
employ the component formalism, with supersymmetry not manifest,
complicating their interpretation.  In this letter, we will study the gap
equation in a superspace formulation.

   A previous treatment of this topic \cite{ClarkLove} concluded that chiral
symmetry does not break unless the Lagrangian explicitly breaks supersymmetry.
We will follow a similar program, but arrive at different conclusions,
arguing that the vanishing of gap equation coefficients in special gauges
does not prohibit chiral symmetry breaking.  We shall separately emphasize
that in general the supersymmetric no-renormalization theorem for the 1PI
effective potential does not apply to a significant class of theories,
and is less restrictive than sometimes assumed elsewhere in the literature.

   The simplest gauge theory, Abelian supersymmetric QED, suffices to
illustrate the main features of the gap equation analysis.  Non-abelian
generalizations, whatever their behaviors in alternative analyses,
introduce no essential complications in the one-loop truncated gap equation.
We thus begin with the SQED action
\begin{eqnarray} \label{SQEDaction}
\int d^4 x \ d^4 \theta \
 {1 \over 8}\, \Big\{ \, V D \bar D \bar D D V \ - \
 {1\over\xi}\, V D D \bar D \bar D V \, \Big\} \nonumber \\
 \ + \ \int d^4 x \ d^4 \theta \
 \Big\{ \, \Phi_+^\dagger \ e^{\, 2 e V } \, \Phi_+ \
 + \ \Phi_-^\dagger \ e^{ - 2 e V } \, \Phi_- \, \Big\} \\
 \ + \ \int d^4 x \ d^2 \theta \
 m\, \Phi_+ \Phi_- \ + \
 \int d^4 x \ d^2 \bar \theta \
 m\, \Phi_+^\dagger \Phi_-^\dagger \nonumber
\end{eqnarray}
using standard conventions \cite{WessBagger} for supersymmetric operator
normalizations and identities.
The chiral field $\Phi_+$ contains a left handed fermion and its scalar
partner, while $\Phi_-$ contains the left-handed anti-fermion and another
scalar partner.  If the bare mass $m$ vanishes, the theory has
a chiral symmetry, separate phase rotations on $\Phi_+$ and $\Phi_-$
(or $SU(N) \times SU(N)$ rotations, in the $N$ flavor case).

\begin{figure}[hbt]
\hfil{
 \epsfysize = 12em
 \epsfbox{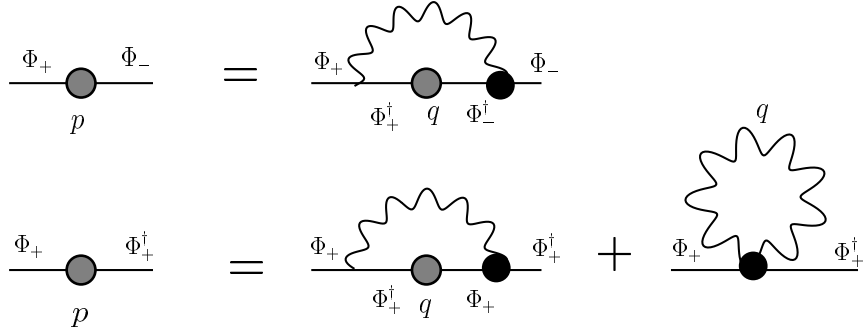}
}\hfil
\caption{
 Contributions to mass and wave function gap equations for the matter
 superfield.  Solid lines with blobs are matter superfield propagators,
 solutions of the gap equation through which the indicated momentum flows.
 Wavy lines are gauge superfield propagators, and black vertex blobs are
 corrected vertices.
} \label{fig1}
\end{figure}

   The gauge invariant condensate $\langle\, \int{d^2\theta}\ \Phi_+(z)\,
 \Phi_-(z)\, \rangle$ varies under chiral transformations, and provides a
suitable order parameter for chiral symmetry breaking.  Since this is the
zero-distance limit of (the fermion part of) the matter field chiral-chiral
propagator, we can shift our attention to the matter field propagators
(quantum corrected beyond finite order in perturbation theory).
We write the full GRS chiral-chiral propagator (for Euclidean momenta) as
\begin{equation}
i \ { -1 \over p^2_{_E} \ Z^2 ( p^2_{_E} ) \ + \
\Sigma^2 ( p^2_{_E} ) } \ \
{ \Sigma^2 ( p^2_{_E} ) \ D^2 \over - 4 \, p^2_{_E} } \ \
\delta^4 ( \theta_1 - \theta_2 )
\end{equation}
and the chiral-antichiral propagator as
\begin{eqnarray}
i \ { -1 \over p^2_{_E} \ Z^2 ( p^2_{_E} ) \ + \
\Sigma^2 ( p^2_{_E} ) } \ \
Z^2 ( p^2_{_E} ) \ \
\delta^4 ( \theta_1 - \theta_2 )\ .
\end{eqnarray}
Coupled Schwinger-Dyson equations relate the inverse of these full propagators
to the inverse of the corresponding bare propagators, together with the
fully resummed one loop corrections shown in Fig.~\ref{fig1}.
Our superspace gap equations will be similar to those of Clark and Love
\cite{ClarkLove}, up to some sign differences and our retention of the
order $e^2$ seagull graph contributing to wave function renormalization.
We will similarly use vertex functions related to a single wave function
factor \cite{ClarkLove,ALM}, in accordance with the Ward identity, rather
than attempting to solve the infinite set of coupled higher-point
equations.\footnote{
 Although the Schwinger-Dyson equations insert only one corrected vertex
 (Fig.~\ref{fig1}), linearizing in $\Sigma$ would correspond to inserting
 two corrected vertex factors \cite{ALM}.
}

   To obtain the gap equations, we must evaluate the diagrams.  After Wick
rotation, the mass renormalization diagram (Fig.~\ref{fig1}a) has the value
\begin{eqnarray}
i \int { d^4 q_{_E} \over ( 2 \pi )^4 } \
4 e^2 \ \Gamma_3 ( p , -q ) \ ( 1 - \xi ) \
{ \Sigma ( q^2_{_E} ) \over
q_{_E}^2 \ Z^2 ( q^2_{_E} ) \ + \ \Sigma^2 ( q^2_{_E} ) } \
{ 1 \over ( p_{_E} - q_{_E} )^4 } \nonumber \\
\int d^4 \theta \ \Phi_- ( p_{_E} , \theta , \bar \theta )
\ { D^2 \over 4 } \, \Phi_+ ( - p_{_E} , \theta , \bar \theta )\ .
\end{eqnarray}
For the chiral superfield wave function renormalization, there are two
graphs (Fig.~\ref{fig1}b) at lowest order in $e^2$.
The gauge emission/absorption diagram is
\begin{eqnarray}
i \int { d^4 q_{_E} \over ( 2 \pi )^4 } \
4 e^2 \ \Gamma_3 ( p , -q ) \
{ Z ( q^2_{_E} ) \over
q_{_E}^2 \ Z^2 ( q^2_{_E} ) \ + \ \Sigma^2 ( q^2_{_E} ) } \
{ 1 \over ( p_{_E} - q_{_E} )^4 } \
\big[ \ p_{_E} \cdot q_{_E} \ - \ { \xi \over 2  } \
( p_{_E}^2 + q_{_E}^2 ) \ \big] \nonumber \\
\times \int d^4 \theta \ \Phi_+^\dagger ( p_{_E} , \theta , \bar \theta )
\ \Phi_+ ( - p_{_E} , \theta , \bar \theta )
\end{eqnarray}
and the seagull gauge loop is
\begin{equation}
- i \int { d^4 q_{_E} \over ( 2 \pi )^4 } \
2 e^2 \ \Gamma_4 ( p , -p , -q ) \ ( 1 - \xi ) \
{ 1 \over q_{_E}^4 } \
\int d^4 \theta \ \Phi_+^\dagger ( p_{_E} , \theta , \bar \theta )
\ \Phi_+ ( - p_{_E} , \theta , \bar \theta ) \ .
\end{equation}
Here $\Gamma_3$ and $\Gamma_4$ are the fully corrected three and four
point vertices.  An approximate form relying on the Ward identities sets
them to $Z(M^2)$, where $M^2$ is the maximum of their squared (Euclidean)
arguments.  The gap equations are then
\begin{equation} \label{massgap}
\Sigma (p^2_{_E}) \ = \  m \ + \
p_{_E}^2 \ 4 e^2 \ ( 1 - \xi )
\int { d^4 q_{_E} \over ( 2 \pi )^4 } \
{ Z ( M^2 ) \ \Sigma ( q^2_{_E} ) \over
q_{_E}^2 \ Z^2 ( q^2_{_E} ) \ + \ \Sigma^2 ( q^2_{_E} ) } \
{ 1 \over ( p_{_E} - q_{_E} )^4 }
\end{equation}
and
\begin{eqnarray} \label{Zgap}
Z ( q^2_{_E} )  \ = \ 1 \ + \
4 e^2 \int { d^4 q_{_E} \over ( 2 \pi )^4 } \
{ Z ( M^2 ) \ Z ( q^2_{_E} ) \over
q_{_E}^2 \ Z^2 ( q^2_{_E} ) \ + \ \Sigma^2 ( q^2_{_E} ) } \
{ 1 \over ( p_{_E} - q_{_E} )^4 } \
\big[ \ p_{_E} \cdot q_{_E} \ - \ { \xi \over 2  } \
( p_{_E}^2 + q_{_E}^2 ) \ \big] \ \nonumber \\
- \ 2 e^2 \ ( 1 - \xi )
\int { d^4 q_{_E} \over ( 2 \pi )^4 } \
{ 1 \over q_{_E}^4 } \ .
\end{eqnarray}
We will not explicitly renormalize the ultraviolet log in Eq.~(\ref{Zgap});
this could be straightforwardly accomplished by changing the leading term
of unity to a scheme-dependent constant $Z_\phi\, (\mu)$ corresponding to
fields $\Phi_\pm$ renormalized at some scale $\mu$.

   The mass gap equation (\ref{massgap}) involves a nonvanishing integral,
but there is no conflict with the supersymmetric no-renormalization theorem.
The latter is sometimes misconstrued in the literature as prohibiting all
corrections to the one-particle-irreducible (as opposed to Wilsonian)
perturbative effective superpotential.  Stated precisely, however, it
merely restricts the possible form of such corrections \cite{West-book}.
(Power counting still indicates the absence of ultraviolet divergences
\cite{West-book,powercount}.)  Substituting $\Sigma = m$ and $Z = 1$ in
the integral of Eq.~(\ref{massgap}) reproduces the one loop contribution
\cite{QEDm} to the one-particle-irreducible self energy.
This correction is not however strictly a superpotential contribution:
the factor of external momentum squared corresponds to a derivative operator,
although not even that was required by the no-renormalization theorem.
For instance, the massless Wess-Zumino model, with or without a Yang-Mills
sector, has perturbative mass corrections without prefactors of external
momenta \cite{nomass}.  Massless particles are necessary for such
non-derivative superpotential contributions, but not sufficient
(as shown by Eq.~\ref{massgap} induced by the massless SQED gauge multiplet).

   The gauge parameter dependence of this result offers another possible
argument for a vanishing dynamical mass.  Clark and Love \cite{ClarkLove}
noted that the gap equation integral is multiplied by zero in Feynman gauge
($\xi=1$), leaving only the trivial solution.  They concluded that SQED in
the gap equation formalism does not break chiral symmetry, unless the
Lagrangian contains explicit supersymmetry breaking.
We should note, however, that the integral evaluated for gap equation
solutions could cancel the vanishing prefactor.  This in fact happens in
non-supersymmetric gauge theory \cite{ALM}, and also in supersymmetric gauge
theory in component formalism \cite{Shamir,ANS}.  The linearized gap equation
(valid in parameter ranges where the dynamical mass is small) has solutions
independent of $\xi$, including the limit of the prefactor approaching zero.
Carrying out the integral in superspace is more complicated, due to
difficulties in regulating infrared divergences (arising, as Clark and Love
discussed, from propagating a supersymmetry-gauge artifact-- the lowest
component of the vector superfield).
The favored infrared regulator prescription \cite{PiguetSibold} gives finite
answers only for gauge invariant Green functions; however, the propagators of
interest here are of course gauge dependent.  Demonstrating the existence of
explicit solutions will require further analysis.

   We conclude that gauge independence arguments, and the no-renormalization
theorem, do not prevent superfield gap equations from yielding chiral
symmetry breaking, even without explicit supersymmetry breaking in the
Lagrangian.  This is reassuring for the corresponding calculations in
component notation, which cannot impose manifest supersymmetry on the
off-shell Green's functions in the gap equation.  Since gauge dependent
infrared contributions are however harder to handle in superspace, the
two formalisms play a complementary role in comparisons of gap equation
techniques with holomorphic/Wilsonian ones.

\vskip 3em

   The authors are grateful to Thomas Appelquist, Thomas Clark, Andrew Cohen,
Nick Evans, Stephen Hsu, Sherwin Love, Mychola Schwetz, Warren Siegel and
Peter West for useful discussions and comments.  This work was supported
in part by the United States Department of Energy under DOE-HEP Grant
DE-FG02-92ER-4074.

\vfil


\baselineskip=1.6pt
\begin{thebibliography}{99}
%

\bibitem{holomorf}
See for example K. Intriligator and N. Seiberg, hep-th/9509066,
Nucl.\ Phys.\ Proc.\ Suppl.\ 45BC, 1 (1996) and references therein.

\bibitem{Shamir}
Y. Shamir, Nucl.\ Phys.\ B352, 469 (1991).

\bibitem{ANS}
T. Appelquist, A. Nyffeler and S.B. Selipsky, Yale preprint YCTP-P12-97,
 in preparation.

\bibitem{ClarkLove} 
T.E. Clark and S.T. Love, Nucl.\ Phys.\ B310, 371 (1988).

\bibitem{WessBagger}
J. Wess and J. Bagger, Supersymmetry and Supergravity, 2nd ed.\
 (Princeton University Press 1992).

\bibitem{ALM}
T. Appelquist, K. Lane and U. Mahanta, Phys.\ Rev.\ Lett.\ 61, 1553 (1988).

\bibitem{West-book}
P. West, Introduction to Supersymmetry and Supergravity, 2nd ed.\
 (World Scientific, 1990).

\bibitem{powercount}
S. Ferrara and O. Piguet, Nucl.\ Phys.\ B93, 261 (1975).

\bibitem{QEDm}
J. Goity, T. Kugo and R.D. Peccei, Phys.\ Rev.\ D29, 2412 (1984).

\bibitem{nomass}
I. Jack, D.R.T. Jones and P. West, Phys.\ Lett.\ B258, 382 (1991),
 and references therein;\\
A. Pickering and P. West, Phys.\ Lett.\ B383, 54 (1996); \\
M.A. Shifman and A.I. Vainshtein, Nucl.\ Phys.\ B359, 571 (1991).

\bibitem{PiguetSibold}
O. Piguet and K. Sibold, Nucl.\ Phys.\ 248, 336 (1984);
 Nucl.\ Phys.\ 249, 396 (1985).

\end{thebibliography}
\end{document}